\documentclass{ws-procs9x6}
\usepackage{graphicx}

\newcommand {\ignore}[1]{}

\def\slash#1{#1\!\!\! /}
\def\Slash#1{#1\!\!\!\! /}
\def\ifmath#1{\relax\ifmmode #1\else $#1$\fi}

\def\ra{\rightarrow}
\def\Eq#1{Eq. (\ref{#1})}

\def\te{{\tilde e}}
\def\tm{{\tilde \mu}}
\def\st{{\tilde \tau}}

\def\rp{$R_p \hspace{-1em}/\;\:$ }

\def\bold#1{\setbox0=\hbox{$#1$} 
     \kern-.025em\copy0\kern-\wd0 
     \kern.05em\copy0\kern-\wd0 
     \kern-.025em\raise.0433em\box0 }
\newcommand{\bemartin}[1]{\begin{equation} \label{(#1)}}
\newcommand{\eemartin}{\end{equation}}
\newcommand{\bamartin}[1]{\begin{eqnarray} \label{(#1)}}
\newcommand{\eamartin}{\end{eqnarray}} 

\newcommand {\chiz} [1] {\tilde{\chi}^{0}_{#1} }

\begin{document}

\title{Testing Neutrino Parameters at Future 
Accelerators\footnote{\uppercase{T}his work is supported by the
  \uppercase{E}uropean \uppercase{C}ommission \uppercase{RTN} network
  \uppercase{HPRN-CT}-2000-00148.}} 

\author{J.~C.~Rom\~ao}

\address{Departamento de F\'{\i}sica, Instituto Superior T\'ecnico\\
Av. Rovisco Pais~1,1049-001 Lisboa, Portugal \\
E-mail: jorge.romao@ist.utl.pt}


\maketitle

\abstracts{
The simplest unified extension of the Minimal Supersymmetric Standard
Model with bilinear R--Parity violation provides a predictive scheme
for neutrino masses which can account for the observed atmospheric and
solar neutrino anomalies.
Despite the smallness of neutrino masses R-parity violation is
observable at present and future high-energy colliders, providing an
unambiguous cross-check of the model.
}

\section{Introduction}
The announcement of high statistics atmospheric neutrino data
by the SuperKamio\-kande collaboration \cite{kamiokande} has
confirmed the deficit of muon neutrinos, especially at small zenith
angles, opening a new era in neutrino physics. 
Although there may be alternative solutions of the atmospheric
neutrino anomaly ~\cite{Gonzalez-Garcia} it is fair to say that
the simplest interpretation of the data is in terms of $\nu_{\mu}$ to
$\nu_{\tau}$ 
flavor oscillations with maximal mixing. This excludes a large mixing
among $\nu_{\tau}$ and $\nu_e$~\cite{kamiokande}, in agreement also
with the CHOOZ reactor data~\cite{chooz}.   On the other hand the
persistent disagreement between solar neutrino data and theoretical
expectations~\cite{BP98} has been a long-standing problem in physics. 
Recent solar neutrino
data~\cite{SNO} are consistent with both
vacuum oscillations and MSW conversions. In the latter case 
the large mixing angle solutions are now
clearly preferred ~\cite{Gonzalez-Garcia}.

Many attempts have appeared in the literature to explain the
data. Here we review recent  results~\cite{numass} obtained in a 
model~\cite{epsrad} which is a simple
extension of the MSSM with bilinear R-parity violation (BRpV). 
This model, despite being a minimal extension of the MSSM, can explain
the solar and atmospheric neutrino data. Its most attractive feature
is that it gives definite predictions for accelerator physics for the
same range of parameters that explain the neutrino data.

\section{Bilinear R-Parity Violation}
\subsection{The Model}

Since BRpV SUSY has been discussed in the literature several times
\cite{numass,epsrad,e3others} we will repeat only the main features 
of the model here.  We will follow the notation of \cite{numass}.
The simplest bilinear \rp model (we call it the \rp MSSM) is
characterized by three additional terms in the superpotential
\begin{equation}
\label{eq:Wpot} 
W = W_{MSSM} + W_{\slash R_P} 
\end{equation} 
where $W_{MSSM}$ is the ordinary superpotential of the MSSM and
\begin{equation}
\label{eq:WRPV} 
W_{\slash R_P} = \epsilon_i \widehat
  L_i\widehat H_u.  
\end{equation} 
These bilinear terms, together with the corresponding terms in the
soft SUSY breaking part of the Lagrangian, 
\begin{equation}
\label{eq:Lsoft}
  {\mathcal L}_{soft} = {\mathcal L}_{soft}^{MSSM} + B_i \epsilon_i {\tilde
    L}_i H_u 
\end{equation} 
define the minimal model, which we will adopt throughout this paper.
The appearance of the lepton number violating terms in Eq.
(\ref{eq:Lsoft}) leads in general to non-zero vacuum expectation
values for the scalar neutrinos $\langle {\tilde \nu}_i \rangle$,
called $v_i$ in the rest of this paper, in addition to the VEVs $v_U$
and $v_D$ of the MSSM Higgs fields $H_u^0$ and $H_d^0$.  Together with
the bilinear parameters $\epsilon_i$ the $v_i$ induce mixing between
various particles which in the MSSM are distinguished (only) by lepton
number (or R--parity).  Mixing between the neutrinos and the
neutralinos of the MSSM generates a non-zero
mass for one specific linear superposition of the three neutrino
flavor states of the model at tree-level while 
1-loop corrections  provide mass for the remaining two neutrino states, 
see~\cite{numass}.

\section{Neutrino Masses and Mixings}

\subsection{Tree Level Neutral Fermion Mass Matrix}

In the basis $\psi^{0T}= 
(-i\lambda',-i\lambda^3,\widetilde{H}_d^1,\widetilde{H}_u^2,
\nu_{e}, \nu_{\mu}, \nu_{\tau} )$ 
the neutral fer\-mions mass terms in the Lagrangian are given by 
\begin{equation}
\mathcal{L}_m=-\frac 12(\psi^0)^T{\bold M}_N\psi^0+h.c.   
\end{equation}
where the neutralino/neutrino mass matrix is 
\begin{equation}
{\bold M}_N=\left[  
\begin{array}{cc}  
\mathcal{M}_{\chi^0}& m^T \cr
m & 0 \cr
\end{array}
\right]
\end{equation}
with
\begin{equation}
\mathcal{M}_{\chi^0}=\left[  
\begin{array}{cccc}  
M_1 & 0 & -\frac 12g^{\prime }v_d & \frac 12g^{\prime }v_u \cr
0 & M_2 & \frac 12gv_d & -\frac 12gv_u \cr
-\frac 12g^{\prime }v_d & \frac 12gv_d & 0 & -\mu  \cr
\frac 12g^{\prime }v_u & -\frac 12gv_u &  -\mu & 0  \cr
\end{array}  
\right] 
\quad ; \quad
m=\left[  
\begin{array}{c}
a_1 \cr
a_2 \cr
a_3 
\end{array}  
\right] 
\end{equation}
where $a_i=(-\frac 12g^{\prime }v_i, \frac 12gv_i, 0,\epsilon_i)$. 
This neutralino/neutrino mass matrix is diagonalized by 
\begin{equation}
\mathcal{ N}^*{\bold M}_N\mathcal{N}^{-1}={\rm diag}(m_{\chi^0_1},m_{\chi^0_2}, 
m_{\chi^0_3},m_{\chi^0_4},m_{\nu_1},m_{\nu_2},m_{\nu_3}) 
\label{eq:NeuMdiag} 
\end{equation}

\subsection{Approximate Diagonalization at Tree Level }

If the \rp parameters are small, then we can block-diagonalize 
${\bold M}_N$ approximately to the form 
diag($m_{eff},\mathcal{M}_{\chi^0}$)
\begin{equation}
m_{eff} = - m \cdot \mathcal{M}_{\chi^0}^{-1} m^T = 
\frac{M_1 g^2 + M_2 {g'}^2}{4\, det(\mathcal{M}_{\chi^0})} 
\left(\begin{array}{ccc}
\Lambda_e^2 & \Lambda_e \Lambda_\mu
& \Lambda_e \Lambda_\tau \\
\Lambda_e \Lambda_\mu & \Lambda_\mu^2
& \Lambda_\mu \Lambda_\tau \\
\Lambda_e \Lambda_\tau & \Lambda_\mu \Lambda_\tau & \Lambda_\tau^2
\end{array}\right).
\end{equation}
The matrices $N$ and $V_{\nu}$ diagonalize 
$\mathcal{M}_{\chi^0}$ and $m_{eff}$ 
\begin{equation}
N^{*}\mathcal{M}_{\chi^0} N^{\dagger} = diag(m_{\chi^0_i})
\quad ; \quad
V_{\nu}^T m_{eff} V_{\nu} = diag(0,0,m_{\nu}),
\end{equation}
where 
\begin{equation}
m_{\nu} = Tr(m_{eff}) = 
\frac{M_1 g^2 + M_2 {g'}^2}{4\, det(\mathcal{M}_{\chi^0})} 
|{\vec \Lambda}|^2.
\end{equation}

\subsection{Approximate Formulas for 1--Loop}

\subsubsection{The masses}

Looking at the numerical results~\cite{numass} we found that the most
important contribution came from the bottom-sbottom loop. To gain an
analytical understanding of the results we expanded the exact results
in the small $\Slash{R}_P$ parameters. The result is~\cite{HPVR2002}
\begin{equation}
  \label{eq:approx_mass}
M_{\nu}\simeq c_0 \left(\matrix{
\Lambda_1 \Lambda_1&\Lambda_1 \Lambda_2&\Lambda_1 \Lambda_3\cr
\Lambda_2 \Lambda_1&\Lambda_2 \Lambda_2&\Lambda_2 \Lambda_3\cr
\Lambda_3 \Lambda_1&\Lambda_3 \Lambda_2&\Lambda_3 \Lambda_3}\right)
+c_1 \left(\matrix{
\epsilon_1 \epsilon_1&\epsilon_1 \epsilon_2&\epsilon_1 \epsilon_3\cr
\epsilon_2 \epsilon_1&\epsilon_2 \epsilon_2&\epsilon_2 \epsilon_3\cr
\epsilon_3 \epsilon_1&\epsilon_3 \epsilon_2&\epsilon_3 \epsilon_3}\right)
\end{equation}
where
\begin{equation}
c_0=\frac{M_1\, g^2 +M_2\, g'^2}{4\, \det(M_{\chi^0})}
\quad ; \quad
c_1=\frac{3}{16\pi^2}\, m_b\, \sin 2 \theta_b \, \frac{h^2_b}{\mu^2}\,
\log \frac{m^2_{\tilde{b_2}}}{m^2_{\tilde{b_1}}}
\end{equation}

\noindent
Diagonalization of the mass matrix gives~\cite{HPVR2002}
\begin{eqnarray}
\label{mneu1}
m_{\nu_1}&=& 0 \\[-2mm]
\label{mneu2}
m_{\nu_2}&\simeq& \frac{3}{16\pi^2}\, m_b\, \sin 2 \theta_b \,
\frac{h^2_b}{\mu^2} \ \log \frac{m^2_{\tilde{b_2}}}{m^2_{\tilde{b_1}}}\
 \frac{(\vec \epsilon \times \vec
  \Lambda)^2}{|\vec \Lambda|^2}\\[-2mm]
\label{mneu3}
m_{\nu_3}&\simeq& \frac{M_1\, g^2 +M_2\, g'^2}{4\, \det(M_{\chi^0})}\
|\vec \Lambda|^2
\end{eqnarray}
The formula for $m_{\nu_3}$ is the tree-level formula that we used to
fix the scale of the atmospheric neutrinos by choosing $|\vec
\Lambda|$. Details of the derivation  can
be found in Ref.~\cite{HPVR2002} where the second most important
contribution, coming from the loop with charged Higgs/charged leptons, is also
discussed.

\subsubsection{The mixings}

The atmospheric angle is easily obtained in terms of the ratio
$\Lambda_2/\Lambda_3$. For the solar angle in the same approximation
we also get a simple formula~\cite{HPVR2002},
\begin{equation}
\tan^2 \theta_{sol} = \frac{2 \epsilon^2}{(\epsilon_2+\epsilon_3)^2}
\end{equation}
that is also in very good agreement with the exact result.

\section{Results for the Solar and Atmospheric Neutrinos}

\subsection{The masses}

The BRpV model produces a hierarchical mass spectrum for almost all
choices of parameters. The largest mass can be estimated by the tree
level value using Eq.~(\ref{mneu3}). 
Correct $\Delta m^2_{atm}$ can be easily obtained by an appropriate
choice of $| \vec \Lambda|$. The mass scale for the solar neutrinos is
generated at 1--loop level and therefore depends in a complicated way
in the model parameters. However, in most cases the result of
Eq.~(\ref{mneu2}) is a good approximation and there is no problem in
having both $\Delta m^2_{atm}$ and $\Delta m^2_{solar}$ set to the
correct scales.

\subsection{The mixings}

\begin{figure}[htb]
\begin{tabular}{cc}
\includegraphics[width=54mm]{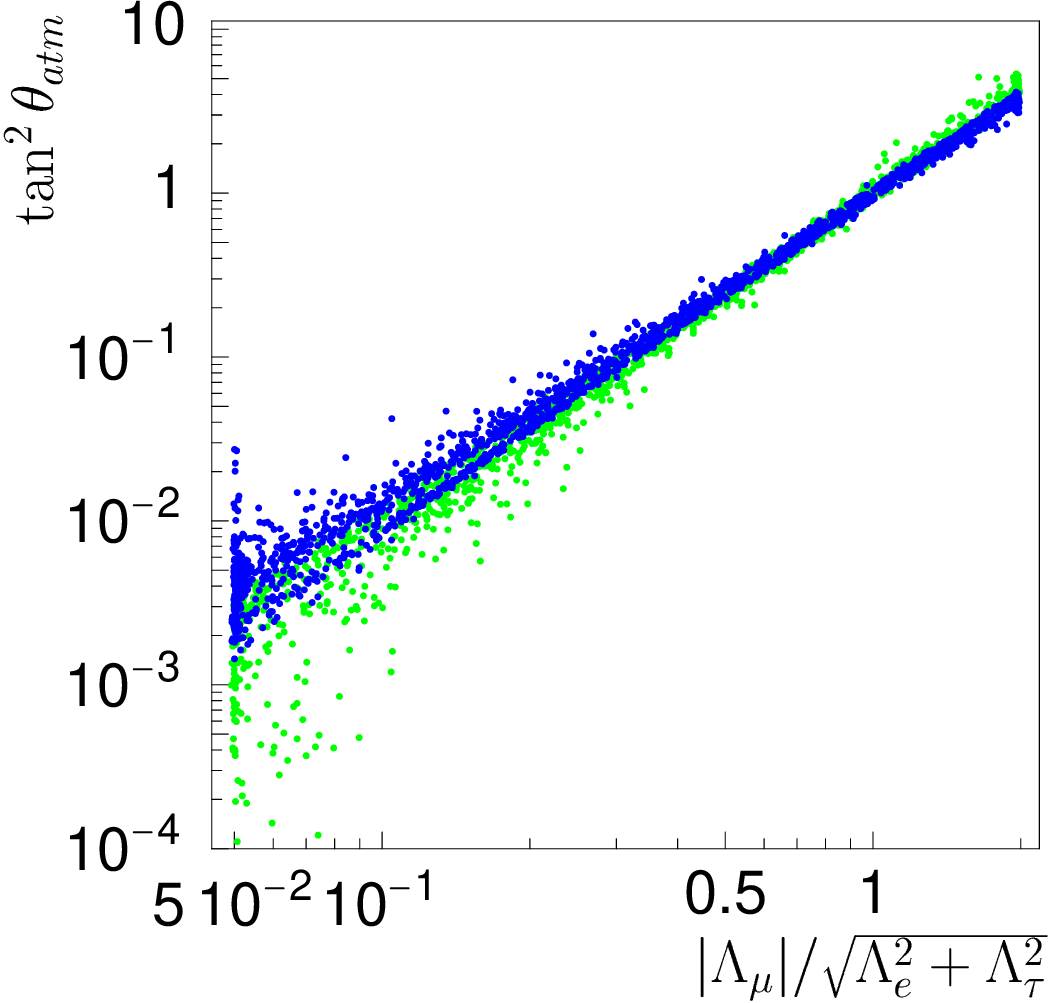}
&
\hskip -1mm
\includegraphics[width=54mm]{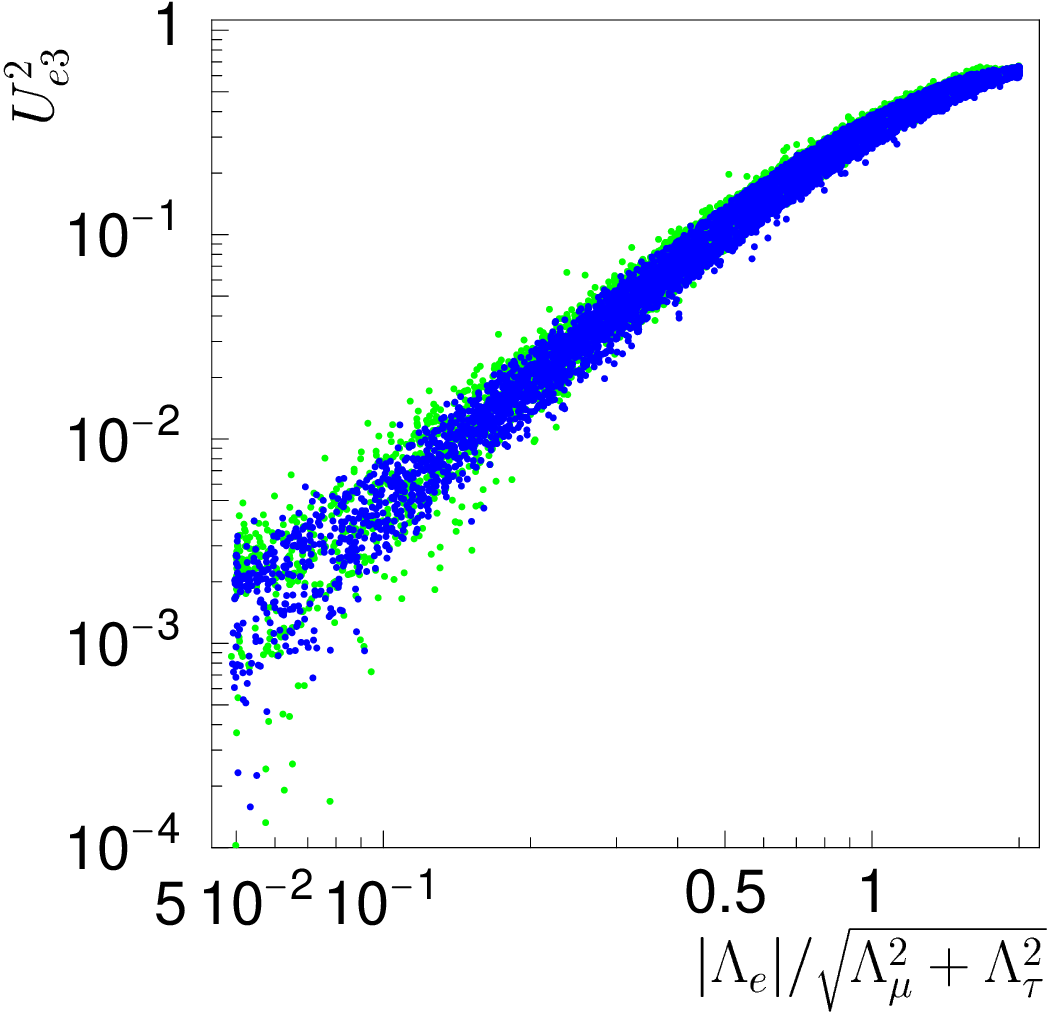}
\end{tabular}
\vspace{-3mm}
\caption{\small 
a) Atmospheric angle as a function of
  $|\Lambda_{\mu}|/\sqrt{\Lambda^2_{e}+\Lambda^2_{\tau}}$. 
  b) $U^2_{e3}$ as a function of
  $|\Lambda_e|/\sqrt{\Lambda^2_{\mu}+\Lambda^2_{\tau}}$.  
}
\label{corfu_fig2}
\end{figure}

Now we turn to the discussion of the mixing angles. We have found that if 
$\epsilon^2/|\vec \Lambda| \ll 100$ then the 1--loop corrections are
not larger than the tree level results and the flavor composition of
the 3rd mass eigenstate is approximately given by
\begin{equation}
U_{\alpha 3}\approx\Lambda_{\alpha}/|\vec \Lambda |
\end{equation}
As the atmospheric and reactor neutrino data tell us that
$\nu_{\mu}\ra \nu_{\tau}$ oscillations are preferred over 
$\nu_{\mu}\ra \nu_e$, we conclude that  
\begin{equation}
\Lambda_e \ll \Lambda_{\mu} \simeq \Lambda_{\tau}
\end{equation}
are required for BRpV to fit the data. This is sown in
Fig.~\ref{corfu_fig2} a). We cannot get so easily maximal mixing for
solar neutrinos, because in this case $U_{e 3}$ would be too large
contradicting the CHOOZ result as shown in Fig.~\ref{corfu_fig2} b).

\begin{figure}[htb]
\begin{tabular}{cc}
\includegraphics[width=55mm,height=51mm]{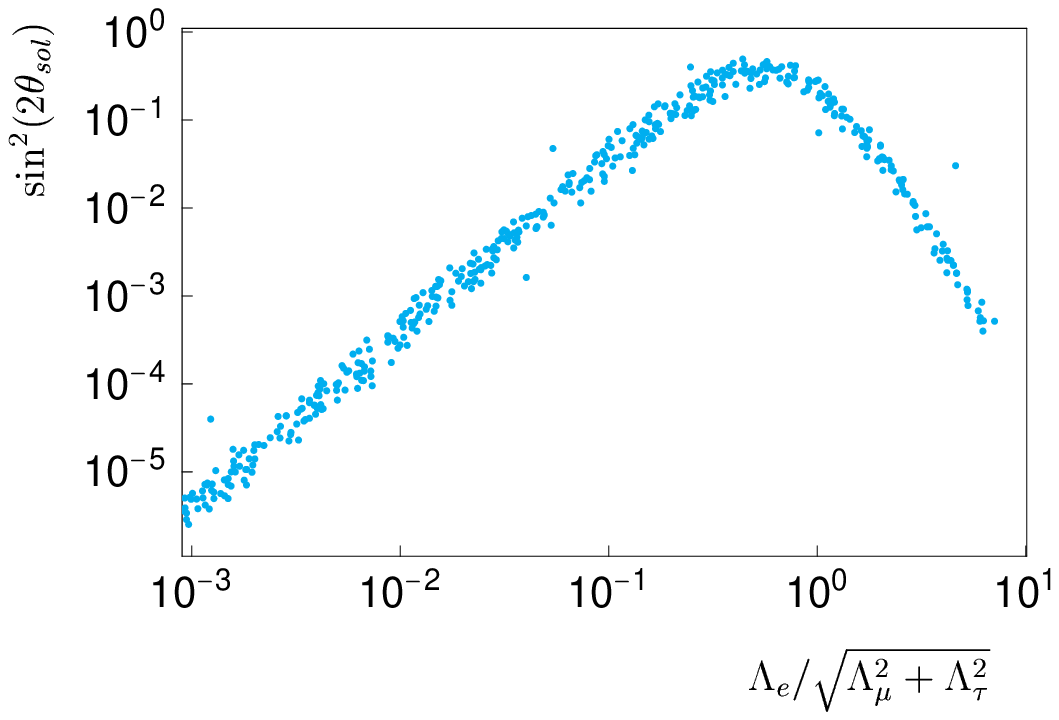}
\vspace{-5mm}
&
\hskip -2mm
\includegraphics[width=53mm,height=50mm,bb=113 479 413 799]{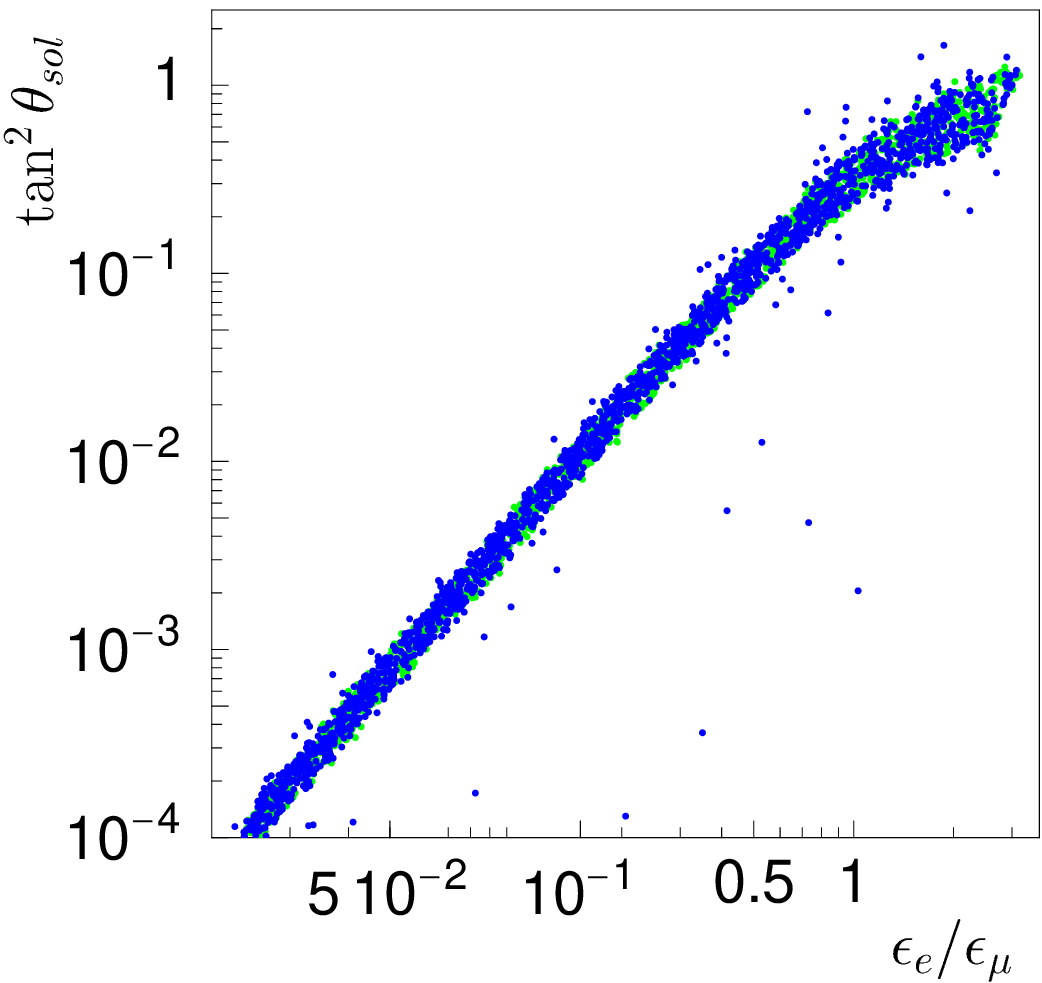}
\end{tabular}
\vspace{-3mm}
\caption{\small
Solar angle as function of: a)
$|\Lambda_e|/\sqrt{\Lambda^2_{\mu}+\Lambda^2_{\tau}}$ \ ;  
b) $\epsilon_e/\epsilon_{\mu}$.
}
\label{corfu_fig3}
\end{figure}

\noindent
We have then two scenarios. In the first one, that we call the
\textit{mSUGRA} case, we have universal boundary conditions of the soft
SUSY breaking terms. In this case we can show~\cite{numass} that 
\begin{equation}
  \label{eq:msugra}
\frac{\epsilon_e}{\epsilon_{\mu}}\simeq \frac{\Lambda_e}{\Lambda_{\mu}}  
\end{equation}
Then from Fig.~\ref{corfu_fig2} b) and the CHOOZ constraint on
$U^2_{e3}$, we conclude that \textit{both} ratios in
Eq.~(\ref{eq:msugra}) have to be small. Then from
Fig.~\ref{corfu_fig3} we conclude that the only possibility is
the small angle mixing solution for the solar neutrino problem. In the
second scenario, which we call the \textit{MSSM} case, we consider
non--universal boundary conditions of the soft SUSY breaking terms. We
have shown that even a very small deviation from universality of the soft
parameters at the GUT scale relaxes this constraint. 
In this case 
\begin{equation}
\frac{\epsilon_e}{\epsilon_{\mu}}\not=\frac{\Lambda_e}{\Lambda_{\mu}}  
\end{equation}
Then we can have at the same time \textbf{small} $U_{e3}^2$ determined by
$\Lambda_e/\Lambda_{\mu}$ as in Fig.~\ref{corfu_fig2} b) 
and \textbf{large} $\tan^2(\theta_{solar})$ determined by
$\epsilon_e/\epsilon_{\mu}$ as in Fig.~\ref{corfu_fig3} b).

\section{Probing Neutrino Mixing via Neutralino Decays}

\begin{figure}[htb]
\begin{tabular}{cc}
\includegraphics[width=54mm,height=49mm]{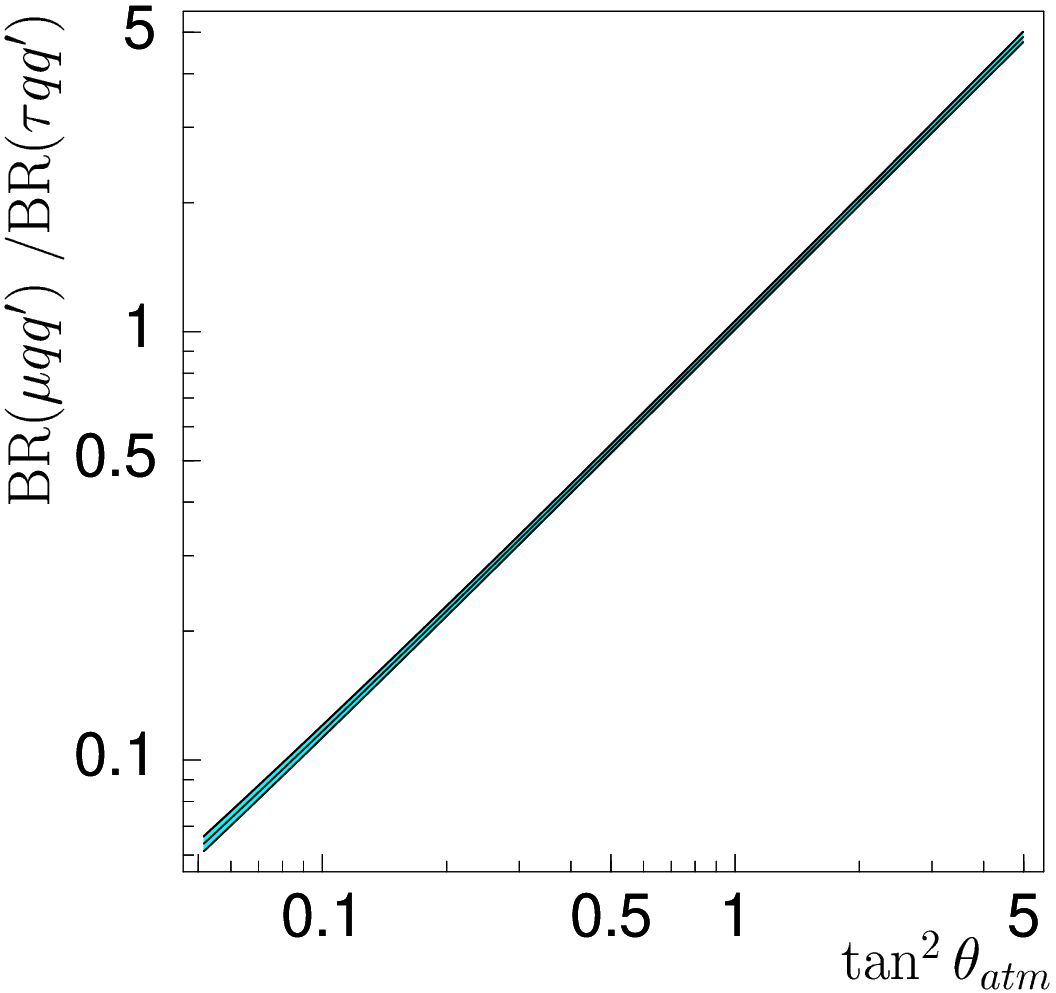}
&
\hskip -1mm
\includegraphics[width=54mm,height=50mm]{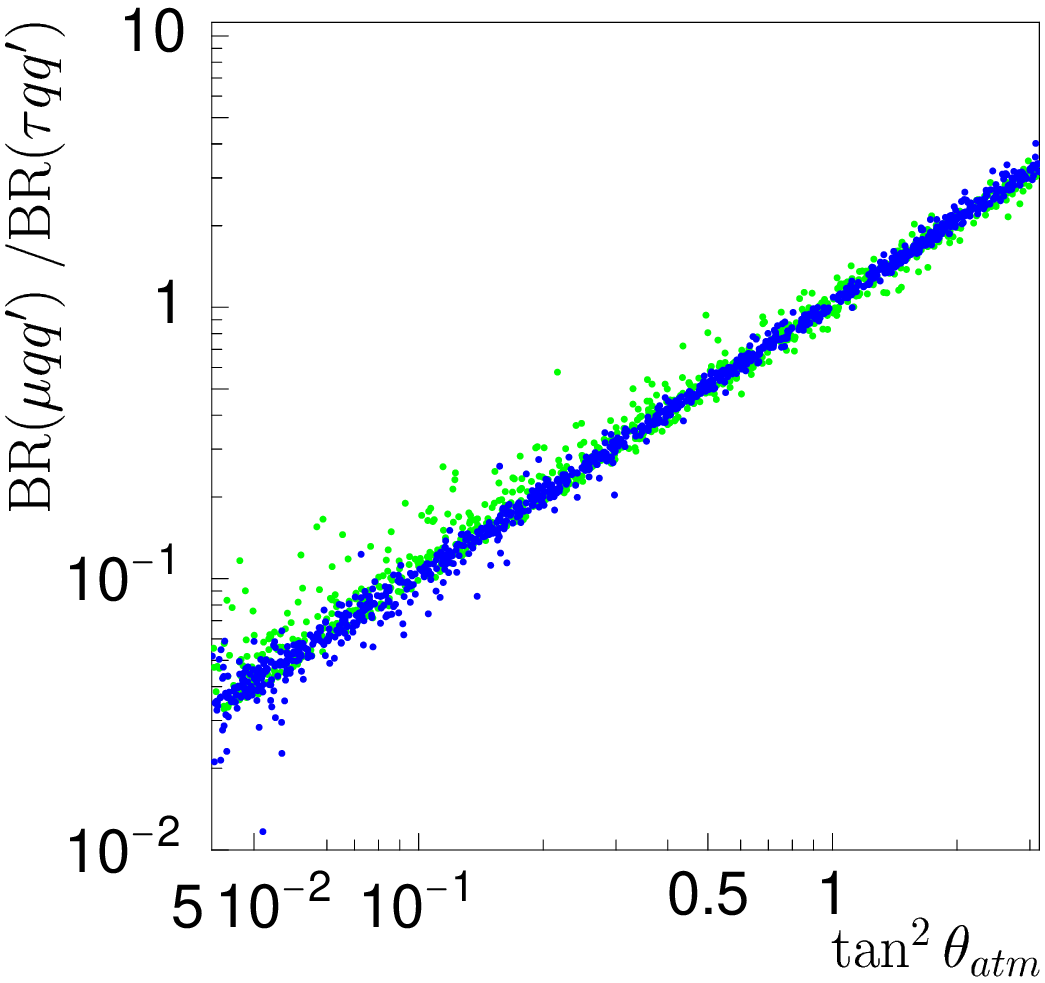}
\end{tabular}
\vspace{-3mm}
\caption{\small
Ratios of semileptonic branching ratios as functions of $\tan
\theta_{atm}$. On the left for random SUSY values and on the right for
the SUSY point of \Eq{eq:SUSYPOINT}
}
\label{faro_fig3}
\end{figure}

If R-parity is broken, the neutralino is unstable and it will decay
through the following channels: $\chiz{1} \to \nu_i \, \nu_j \,
\nu_k,\ \nu_i \, q \, \bar{q}, \ \nu_i \, l^+_j \, l^-_k, \ l^\pm_i \,
q \, \bar{q}', \ \nu_i \, \gamma $. It was shown\footnote{The relation
of the neutrino parameters to the decays of the neutralino has also
been considered in Ref.~\cite{Mukhopadhyaya:1998xj}.} in
Ref.~\cite{Porod:2000hv}, that the neutralino decays well inside the
detectors and that the visible decay channels are quite large. This
was fully discussed in Ref.~\cite{Porod:2000hv} where it was shown
that the ratios $|\Lambda_i/\Lambda_j|$ and
$|\epsilon_i/\epsilon_j|$ were very important in the choice of solutions
for the neutrino mixing angles. What is exciting now, is that these
ratios can be measured in accelerator experiments.  In the left panel
of Fig.~\ref{faro_fig3} we show the ratio of branching ratios
for semileptonic LSP decays into muons and taus: $BR(\chi \to \mu q'
\bar q)/ BR(\chi \to \tau q' \bar q$) as function of $\tan^2
\theta_{atm}$. We can see that there is a strong correlation.  
The spread in this figure can in
fact be explained by the fact that we do not know the SUSY
parameters. This is illustrated in the right panel where
we considered that SUSY was already discovered with the following
values for the parameters,
\begin{equation}
M_2=120\, GeV, \mu=500\, GeV, \tan\beta=5,
 m_0=500\, GeV, A=-500\, GeV
\label{eq:SUSYPOINT}
\end{equation}

\section{Probing Neutrino Mixing via Charged Lepton Decays}

\begin{figure}[htb]
\begin{tabular}{cc}
\includegraphics[width=55mm,height=50mm]{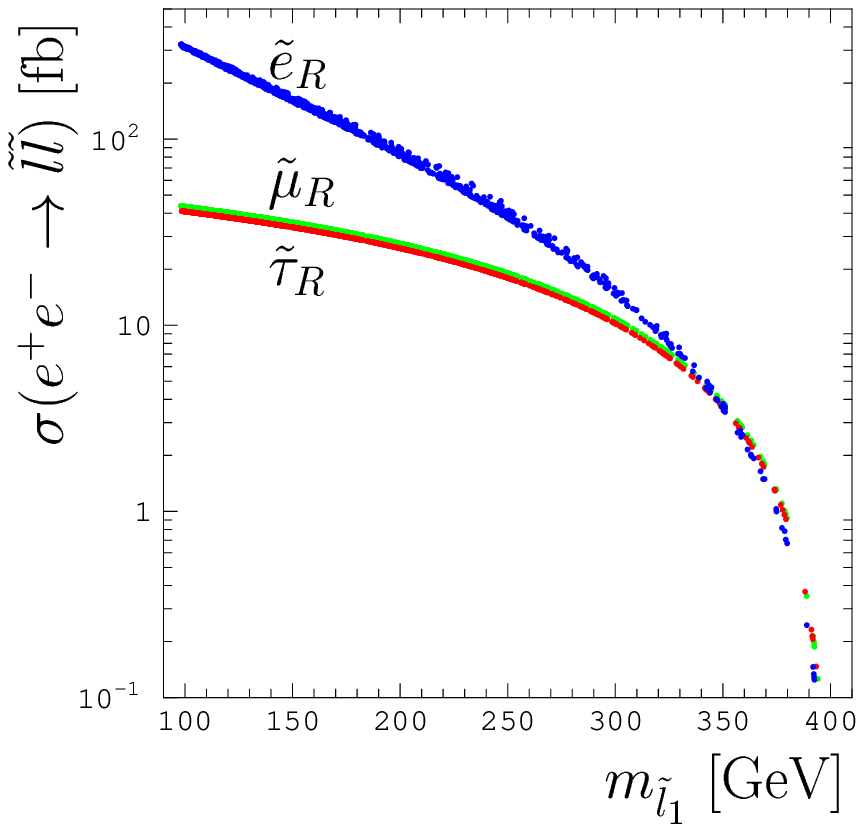}
&
\hskip -3mm
\includegraphics[width=55mm,height=50mm]{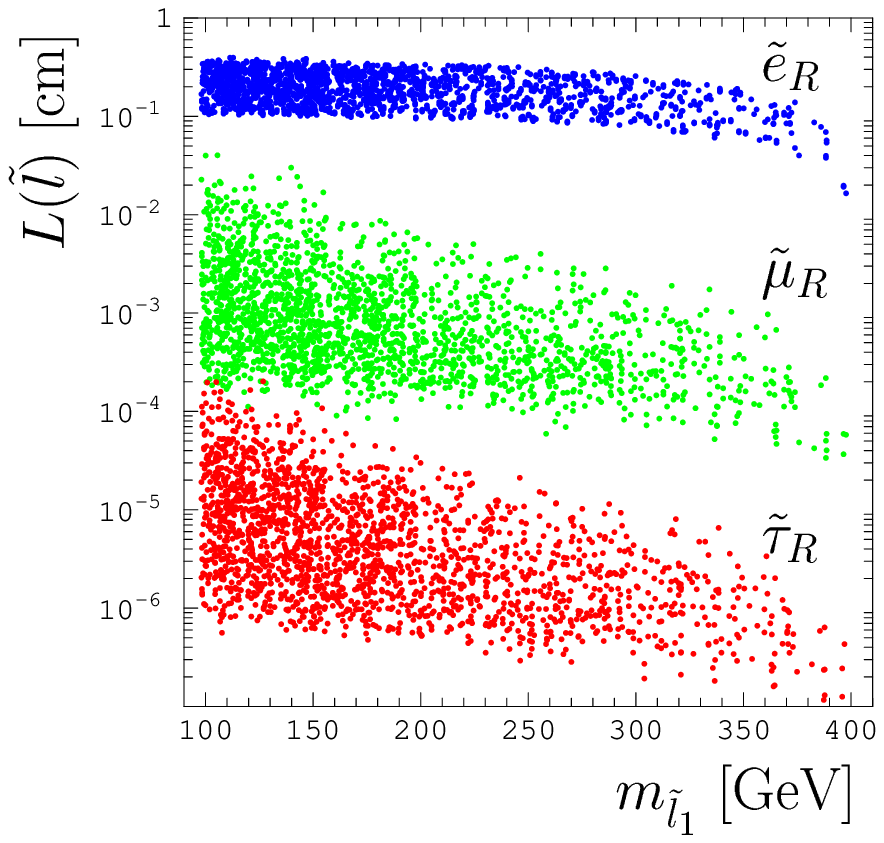}
\end{tabular}
\vspace{-3mm}
\caption{\small
a)$e^+ e^- \to {\tilde l}{\tilde l}$ production 
  cross section  at a Linear Collider $\sqrt{s}= 0.8 \hbox{TeV}$,
b) Charged slepton decay length at a linear collider with $\sqrt{s}=
  0.8 \hbox{TeV}$.
}
\label{faro_fig4}
\end{figure}

\begin{figure}[htb]
\begin{tabular}{cc}
\includegraphics[width=55mm,height=50mm]{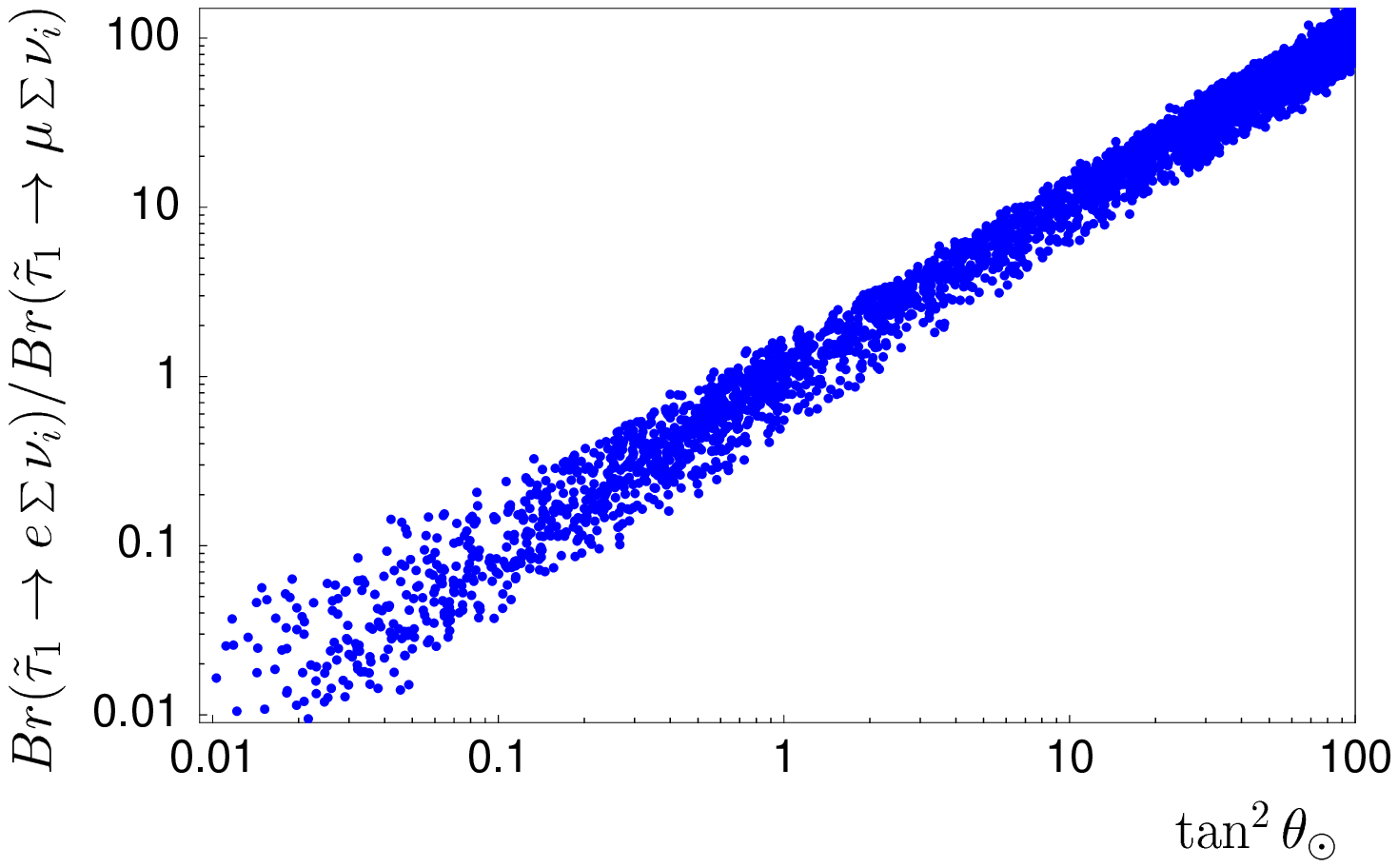}
&
\hskip -3mm
\includegraphics[width=55mm,height=50mm]{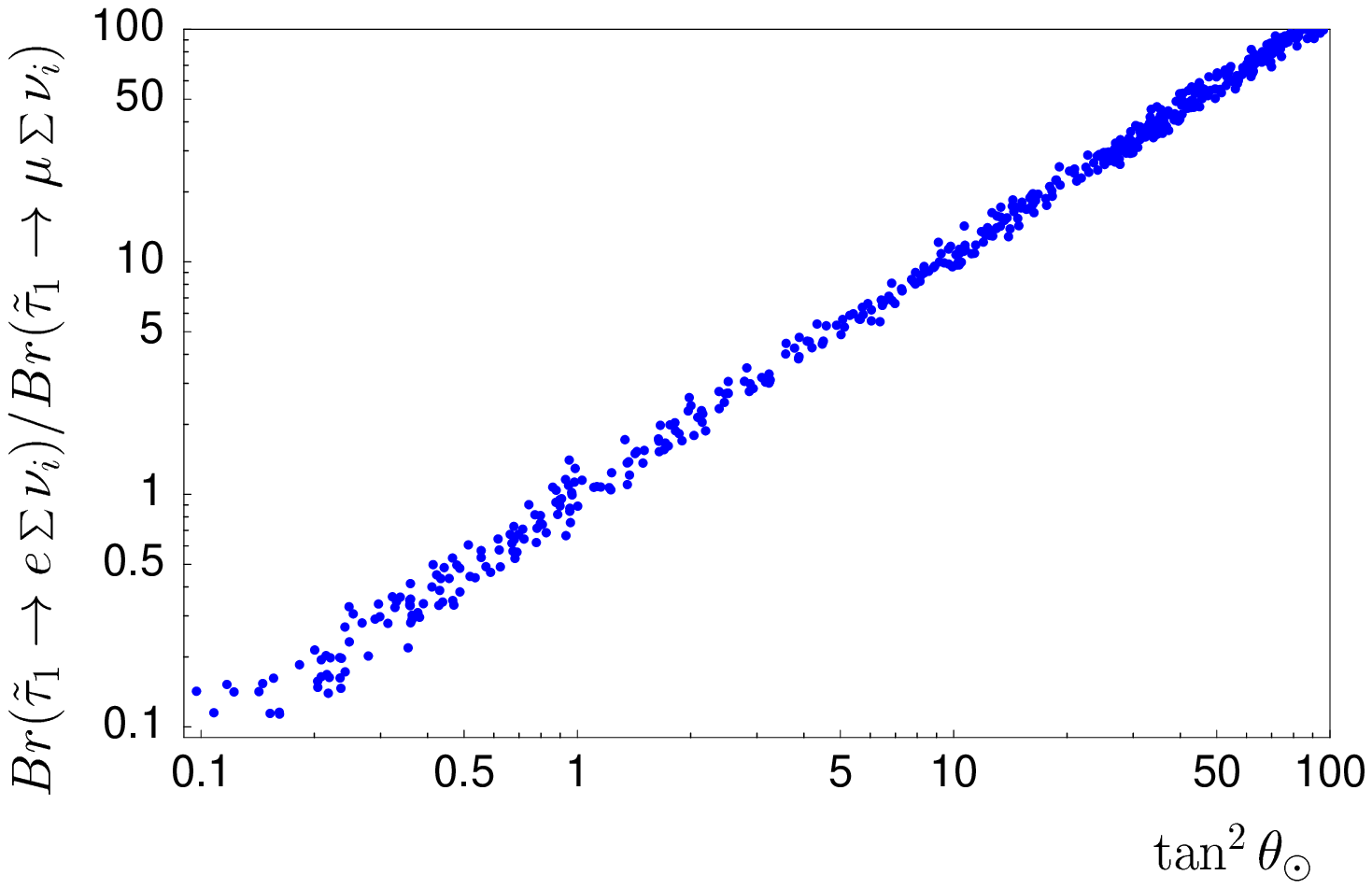}
\end{tabular}
\vspace{-3mm}
\caption{\small
Ratios of branching ratios for scalar tau decays 
  versus $\tan^2\theta_{\odot}$. The left panel shown all data
  points, the right one refers only to data points with
  $\epsilon_2/\epsilon_3$ restricted to the range [0.9,1.1].
}
\label{faro_fig5}
\end{figure}

If R-parity is broken the lightest supersymmetric particle (LSP) will
decay. If the LSP decays then cosmological and astrophysical
constraints on its nature no longer apply. Thus, within R-parity
violating SUSY a priori {\em any} superparticle could be the LSP. 
We have studied~\cite{Hirsch:2002ys} the case where a charged scalar lepton,
most probably the scalar tau, is the LSP. The
production and decays of $\st$, as well as the decays of $\te$ and
$\tm$, and demonstrate that also for the case of charged sleptons as
LSPs neutrino physics leads to definite predictions of various decay
properties. This is shown in Figs.~\ref{faro_fig4} and \ref{faro_fig5}

\section{Conclusions}

The Bilinear R-Parity Violation Model is a simple extension of 
the MSSM that leads to a very rich phenomenology.
We have calculated the one--loop corrected  masses
and mixings for the neutrinos in a completely consistent way,
including the RG equations and correctly minimizing the potential.
We have shown that it is possible to get easily maximal mixing for the
atmospheric neutrinos and both small and large angle MSW.

We emphasize that the LSP decays inside the detectors,
thus leading to a very different phenomenology than the MSSM. 
The LSP can be either the lightest neutralino, like in the MSSM, or a
charged particle, must probably the lightest stau. In both cases we
have shown that ratios of the branching ratios of the LSP can be
correlated with the neutrino parameters.

If the model is to explain solar and atmospheric neutrino problems
many signals will arise at future colliders. These will probe
the neutrino mixing parameters. Thus the model is easily falsifiable!

\end{document}